\documentclass[preprintnumbers,amsmath,amssymb,twocolumn,showpacs]{revtex4}
\usepackage{graphicx}
\usepackage{dcolumn}
\usepackage{bm}

\begin{document}

\title{Interferometric thermometry of a single sub-Doppler cooled atom}

\author{L. Slodi\v{c}ka$^1$, G. H\'etet$^{1}$, N. R\"ock$^1$, S. Gerber$^{1}$, P. Schindler$^1$, M. Kumph$^1$, M. Hennrich$^1$, and R. Blatt$^{1,2}$}

\affiliation{
$^1$ Institute for Experimental Physics, University of Innsbruck, A-6020 Innsbruck, Austria \\
$^2$ Institute for Quantum Optics and Quantum Information of the
Austrian Academy of Sciences, A-6020 Innsbruck, Austria}

\begin{abstract}
Efficient self-interference of single-photons emitted by a
sideband-cooled Barium ion is demonstrated. First, the
technical tools for performing efficient coupling to the quadrupolar
transition of a single $^{138}$Ba$^{+}$ ion are presented. We show efficient
Rabi oscillations of the internal state of the ion using a highly
stabilized 1.76 $\mu m$ fiber laser resonant with the
S$_{1/2}$-D$_{5/2}$ transition. We then show sideband cooling of the ion's
motional modes and use it as a means to enhance the
interference contrast of the ion with its mirror-image to
up to 90\%. Last, we measure the dependence of the self-interference
contrast on the mean phonon number, thereby demonstrating the potential of the set-up for single-atom thermometry close to the motional ground state.  
\end{abstract}
\maketitle

The interaction of a single photon with a single isolated atom at
rest in free space is an important model in
quantum physics.  Systems that enable its precise experimental investigation have become available with optically cooled atoms in harmonic potentials which can be
localized to a few nanometers in position for extended periods of time \cite{Nus05,Dot05,Lei03}.
Over the past decade, remarkable progress has been made in the control of the coupling of single atoms to light with important steps made in the direction of quantum networking \cite{Vol06, Bar09, Pir11}. In particular, single ions in Paul traps \cite{Pau90} are considered today as very promising systems for such applications \cite{Dua10} because of the excellent control of internal and external (motional) states that is available \cite{Lei03}. For instance, a good atom localization is  essential for long distance entanglement of atomic q-bits relying on single photon interference \cite{Cab99}.
On the pure quantum optics side, as one of the most fundamental processes, the quantum properties of resonance fluorescence from single trapped atoms was investigated with a high level of detail \cite{Our11, Ger09, Dar05}. The theoretical grounds for the understanding of the interaction of single atoms  with single photons is very well established and the experimental results are in good agreement with the theoretical expectations.

To allow a different insight into the motional and internal degrees of freedom of single atoms, a boundary condition that modifies its fluorescence characteristics and atomic state was introduced in \cite{Esc01, Dor02} using a single dielectric mirror, thus forming a ``half-cavity". This method was shown to be very useful for investigating novel photon-ion interactions \cite{Dub07, Gla10, Wan12}, quantum-electrodynamic effects \cite{Esc01, Wil03, Het11}, and performing real-time read out and feedback of the ion motional state \cite{Bus06}.
With this set-up, interference of single photons was achieved with visibilities of up to 72\% \cite{Esc01}, mainly limited by atomic motion and aberrations of the optical system.

In this paper, we go a step further in the characterization of the
half-cavity set-up by measuring the self-interference contrast of the emitted single photons for various phonon numbers very close to the motional ground state of a Barium ion. To perform these measurements, we stabilize a fiber laser at 1.76 $\mu$m to a high finesse optical cavity. We show coherent driving of the S$_{1/2}$-D$_{5/2}$ quadrupolar transition and sideband cooling of a single $^{138}$Ba$^{+}$ ion in a linear Paul trap. This experimental results will form the core of the first section of this paper. Then,  the ground-state-cooled ion is used in the half-cavity set-up. The interference contrast is determined as a function of the single-ion mean motional number, and visibilities of up to
90\% are observed with the atom close to the motional ground state. We show that this interferometric method can be used for very precise probing of the temperature of sub-Doppler cooled atoms.

\section{Experimental setup}

\begin{figure}[!h!]
\centerline{\scalebox{0.8}{\includegraphics{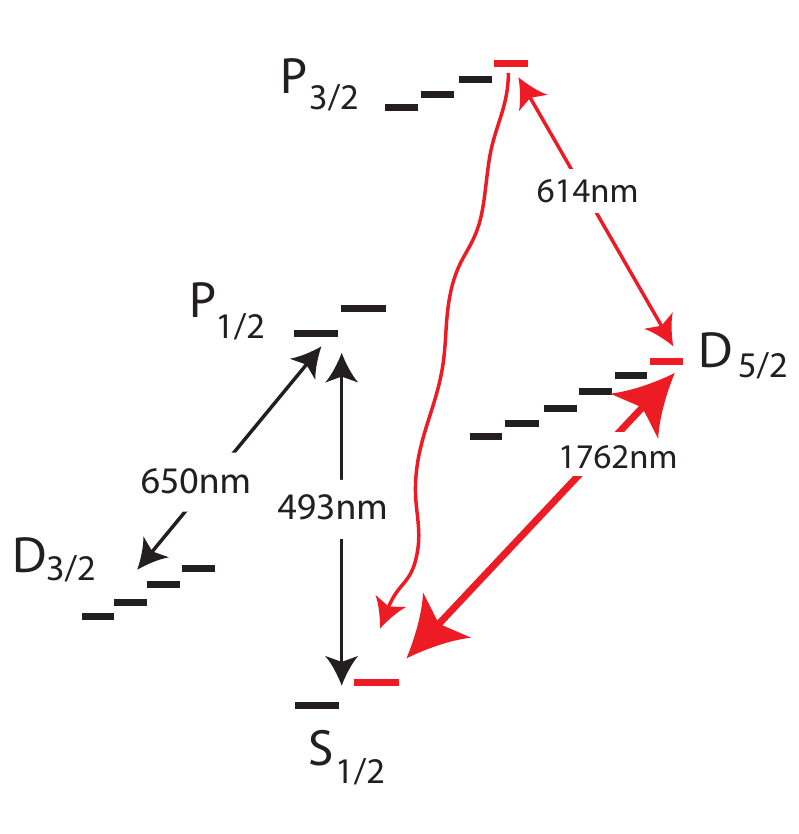}}}
\caption{(Color Online) Level scheme of $^{138}$Ba$^{+}$. The excitation path used for sideband cooling is marked in red.}
\label{lvl}
\end{figure}

We first describe the experimental procedure.
The single $^{138}$Ba$^{+}$ ion is located in a linear Paul trap, with
 radial and the axial frequencies of $(\omega_x,\omega_y,\omega_z)=2\pi
(1.75,1.71,0.85)$\,MHz. The energy level diagram of $^{138}$Ba$^{+}$, is shown
Fig.~\ref{lvl}. The ion is Doppler cooled using a 493 nm laser
and the population is repumped from the
D$_{3/2}$ level using a 650 nm laser beam. The
S$_{1/2}$-P$_{1/2}$-D$_{3/2}$ levels then form a closed cycle.
Using these dipole transitions, the ion is already cooled well
within the Lamb-Dicke regime. In order to reach temperatures close to the ground state of motion, we use sideband-cooling on the quadrupolar transition.

\subsection{Control of the ion quadrupolar transition}

The quadrupolar transition wavelength is $1.76~\mu$m for Barium, a transition which has only recently been used for showing Rabi oscillations with the $^{137}$Ba$^{+}$ isotope \cite{Die10}. Here, we perform Rabi oscillations on $^{138}$Ba$^{+}$, and show efficient sideband cooling of the ion. For sideband cooling, we also use a laser resonant with the $D_{5/2}$ to the $P_{3/2}$ transition at 614 nm (see Fig.~\ref{setup}-b). As in \cite{Die10}, we excite
the ion using a fiber laser
that is actively stabilized to a high finesse cavity.
The laser delivers about 50 mW of power.
An acousto-optic shifter (AO1) is used to shift the laser
frequency by -250 MHz and the remaining 0th order beam is sent to a wavemeter.
The laser is then fiber guided to an optical cavity and to the ion trap.


The optical cavity is made of a 10\,cm long ultra-low expansion
(ULE) spacer (thermal expansion coefficient of less than $10^{-9}/$mK)
and the chamber is actively temperature controlled. As shown in
Fig.~\ref{setup} c), the spacer in fact comprises four cavities
that enable the simultaneous stabilization of four different
lasers. The cavity for the 1.76$~\mu m$ laser is made of high reflectivity mirrors which,
in order to reach a high length stability, are not mounted on
piezo-electric transducers, as is the case for the other dipole lasers.
The cavity is surrounded by an aluminium enclosure and placed in a
stainless steel vacuum can. The vacuum chamber is kept at a
constant pressure of $10^{-7}$ mbar using an ion getter pump. This
provides thermal isolation from the environment and
prevents from pressure changes. For further stabilization, the
vacuum can is enclosed in another aluminium housing to which four
peltier elements are connected. The metal housing is surrounded by
a 5\,cm thick layer of polystyrene for acoustic isolation and the
whole cavity setup is mounted on a vibration isolation platform.
Ring down measurements show a cavity finesse of $1.94 \times
10^{5}$, which corresponds to a linewidth of 7.7~kHz.

\begin{figure}[!h!]
\centerline{\scalebox{0.7}{\includegraphics{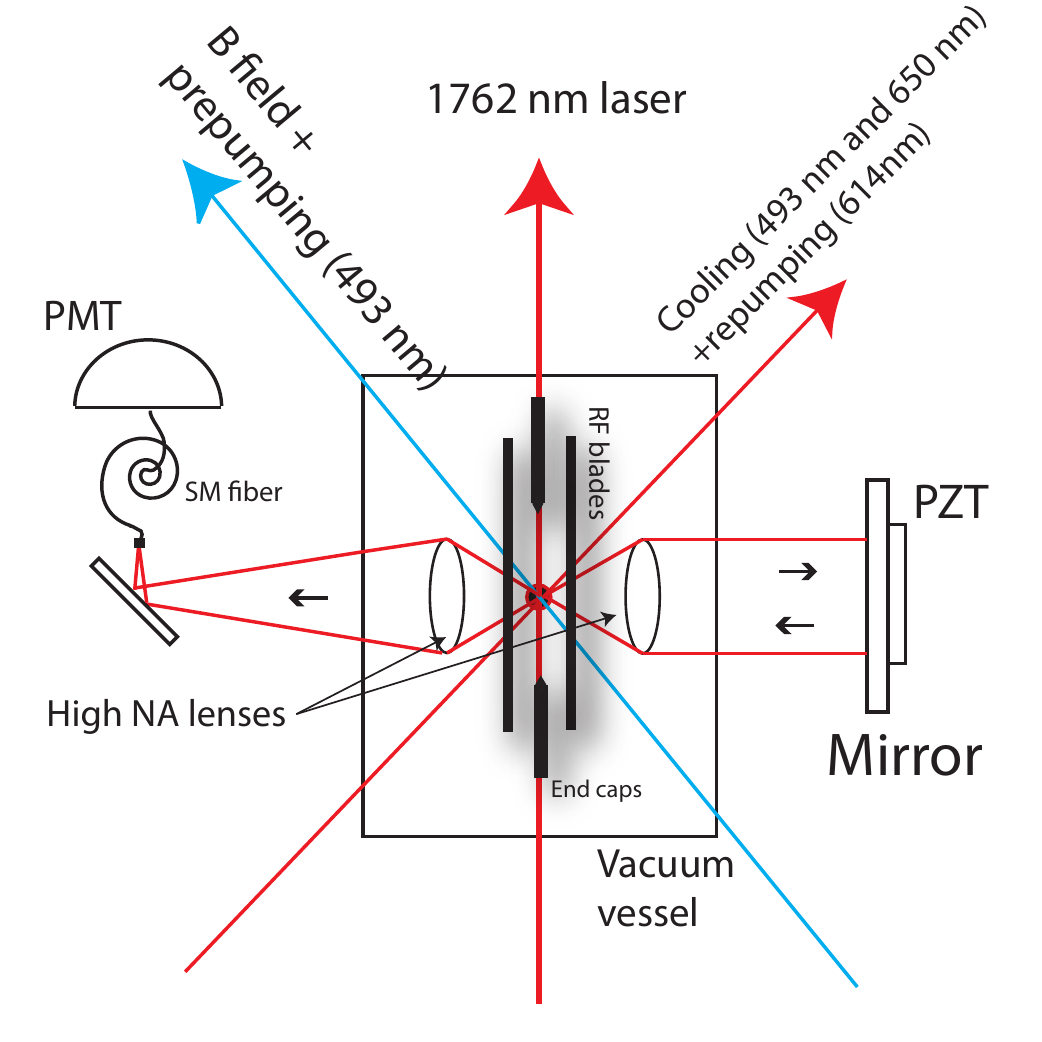}}}
\caption{(Color Online) Schematics of the optical
setup and linear Paul trap. PMT=Photomultiplier tube; SM=single mode; PZT=Piezo-electric Transducer. NA=numerical aperture.}
\label{setup1}
\end{figure}

As shown in Fig.~\ref{setup} c), a Pound-Drever-Hall error signal is derived
using the back-reflected light from the cavity. After going to a
 PI-servo circuit (consisting of a proportional (P) and an integrating (I) part), the low frequency part of the error
signal is fed back to a piezo-electric actuator that controls the
laser frequency. The error signal is also fed back to the AO1 to
compensate for changes of the laser frequency faster than the
piezo response.  Part of the first order diffracted beam from AO1, is then sent to
another acousto-optic modulator (AO2) which can be switched and
frequency controlled using a programmable pulse generator. The
first order diffraction from AO2 is sent through a single mode
polarization maintaining fiber and to an
objective (NA=0.15) close to the trap center. About 4\,mW is then available
to excite the ion after the fiber. By performing a high resolution spectroscopy of
the $S_{1/2}-D_{5/2}$ transition on a single trapped ion, we could
estimate the upper limit for the laser linewidth to be $1 \pm
0.2$\,kHz.

\begin{figure}[!h!]
\centerline{\scalebox{1}{\includegraphics{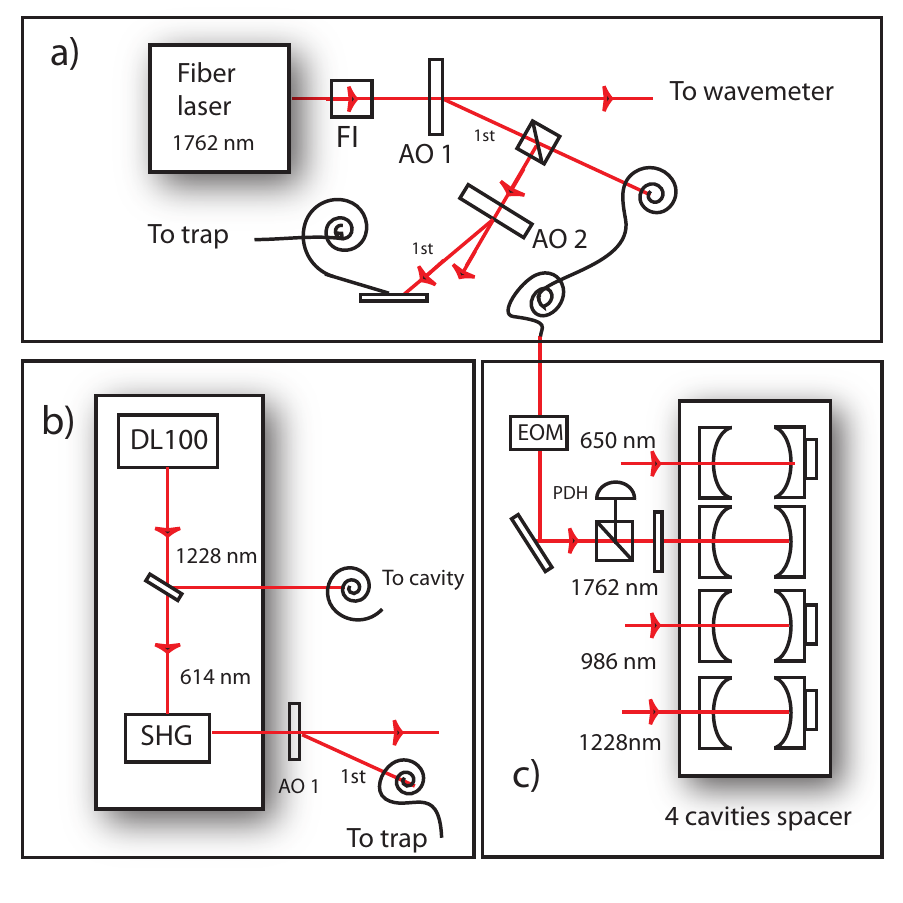}}}
\caption{(Color Online) a) Setup of the 1.76~$\mu$m laser locking. b) 614\,nm
laser light generation. c) Schematics of the four
cavities spacer. SHG=Second Harmonic Generation; FI=Faraday Isolator; AO= Acousto-optic Modulator; PDH=Pound Drever Hall; EOM=Electro-optic modulator}\label{setup}
\end{figure}

In order to perform efficient sideband cooling and
state read-out, we also set up a diode laser system resonant with the $D_{5/2}$ to $P_{3/2}$ transition. As
shown in Fig.~\ref{setup} b), a 1228 nm laser diode is
locked to one of the cavities of the ``4 holes spacer". A
Pound-Drever-Hall error signal is also derived from the cavity but
fed-back to the laser current. The 1228 nm laser then pumps a second
harmonic generator that provides 2\,mW of light at 614 nm. About 100~$\mu W$ are used as a repumper co-propagating with the cooling lasers.

\begin{figure}[!h!]
\centerline{\scalebox{0.8}{\includegraphics{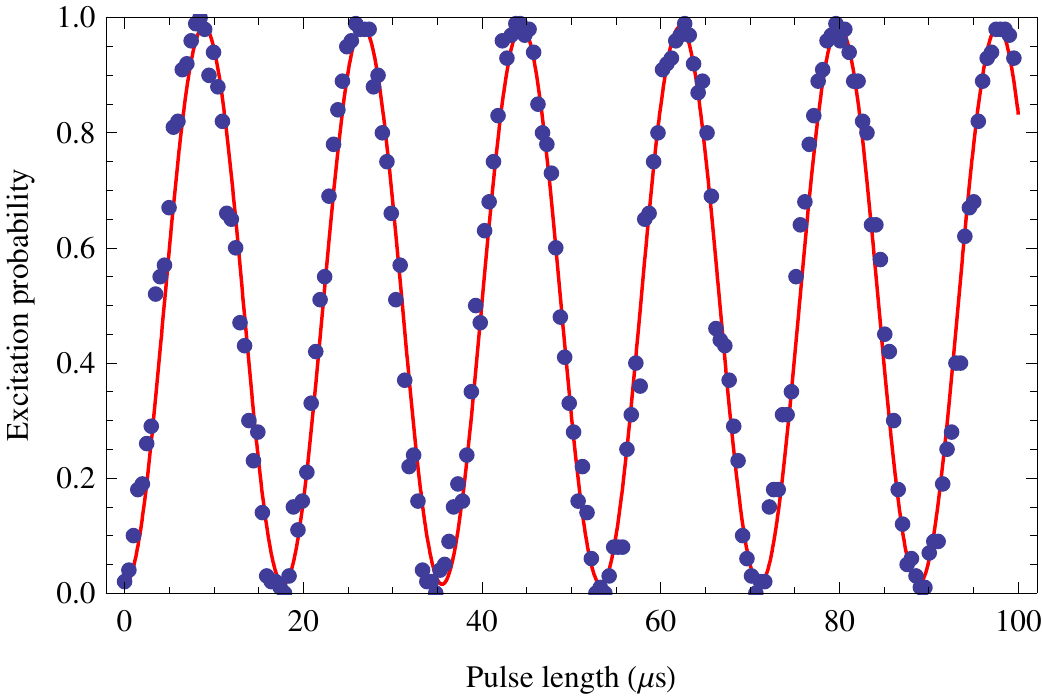}}}
\caption{(Color Online) Probability of exciting the D$_{5/2}$ level as a function of the 1.76~$\mu$m laser pulse length.} \label{flop}
\end{figure}

Fig.~\ref{setup1} shows a top view of the B field direction, trap orientation and laser beams used in these measurements (the 1.76~$\mu$m laser is sent from the bottom but is displayed as being parallel to the trap for simplicity).
To observe coherent oscillations on the quadrupolar transition,
we use a sequence of laser pulses that lasts 2\,ms and repeat this
same sequence 100 times. We first Doppler cool the ion for 1 ms,
and then prepare the ion state in the S$_{1/2}(m_F=-1/2)$ Zeeman sublevel
using a short $\sigma^+$ polarized laser pulse propagating along the magnetic
field. Then a 1.76~$\mu$m laser pulse is generated and
state detection is performed by scattering light from the dipole transitions
(S$_{1/2}$-P$_{1/2}$-D$_{3/2}$). Fluorescence light at 493 nm is collected via a high numerical aperture lens (NA$\approx 0.4$)\cite{Ger09}. By scanning the laser frequency we
can then access the quadrupolar transition spectrum. After
having identified all five carrier transitions, we can for example
measure Rabi oscillations on the
S$_{1/2}(m=-1/2)$-D$_{5/2}(m=-5/2)$ population by scanning the 1.76~$\mu$m
laser pulse duration. A typical plot of such Rabi oscillations is shown in
Fig.~\ref{flop}. At these powers (here the Rabi frequency is about 300 kHz), high contrast of the excitation to the D$_{5/2}$ level are observed for up to $700 \mu$s, limited mostly by the residual motion of the ion and laser intensity and frequency noise.

\subsection{Sideband cooling of Barium}

Doppler cooling only gives mean phonon numbers on the order of 10.
In order to reach a better cooling, we use a
sequence of laser excitations resonant with the red sideband, which can in principle provide a reduction of the
mean phonon number down to the ground state of motion. To enhance
the speed of these cycles of excitation and decay we apply the
614\,nm laser light together with the 1.76~$\mu$m laser. The cooling speed
is then directly proportional to the Lamb-Dicke factor $\eta=k
\cos\theta\sqrt{\hbar/2 m \omega}$, where $k$ is the 1.76~$\mu$m laser wavevector, $\theta$ the angle between the motional oscillation (at
frequency $\omega$) and the 1.76~$\mu m$ laser beam direction, and $m$ the atomic mass \cite{Die89}.
With our laser beam configuration the Lamb-Dicke factor is $\eta=9\times 10^{-3}$
on the axial mode, and 0.011 on the radial modes which means that the cooling time needed
to reach the ground state from a mean phonon number of about
$\overline{n}=10$ is on the order of a few milliseconds.

When performed on the
S$_{1/2}(m=-1/2)$-D$_{5/2}(m=-5/2)$-P$_{3/2}(m=-3/2)$ levels, this
sideband cooling process is a closed cycle. Short pulses of
493 nm $\sigma^+$ light are however applied every 200$\mu s$ on the
ion to depopulate the S$_{1/2}(m=1/2)$ level that might eventually
get populated due to a decay from the P$_{3/2}$ down to the
D$_{3/2}$ level and to the other D$_{5/2}$ Zeeman sub-levels. A
1.76~$\mu$m laser detection pulse is finally scanned across the red and blue
sidebands to estimate the mean photon number after sideband
cooling.

\begin{figure}[!h!]
\centerline{\scalebox{0.7}{\includegraphics{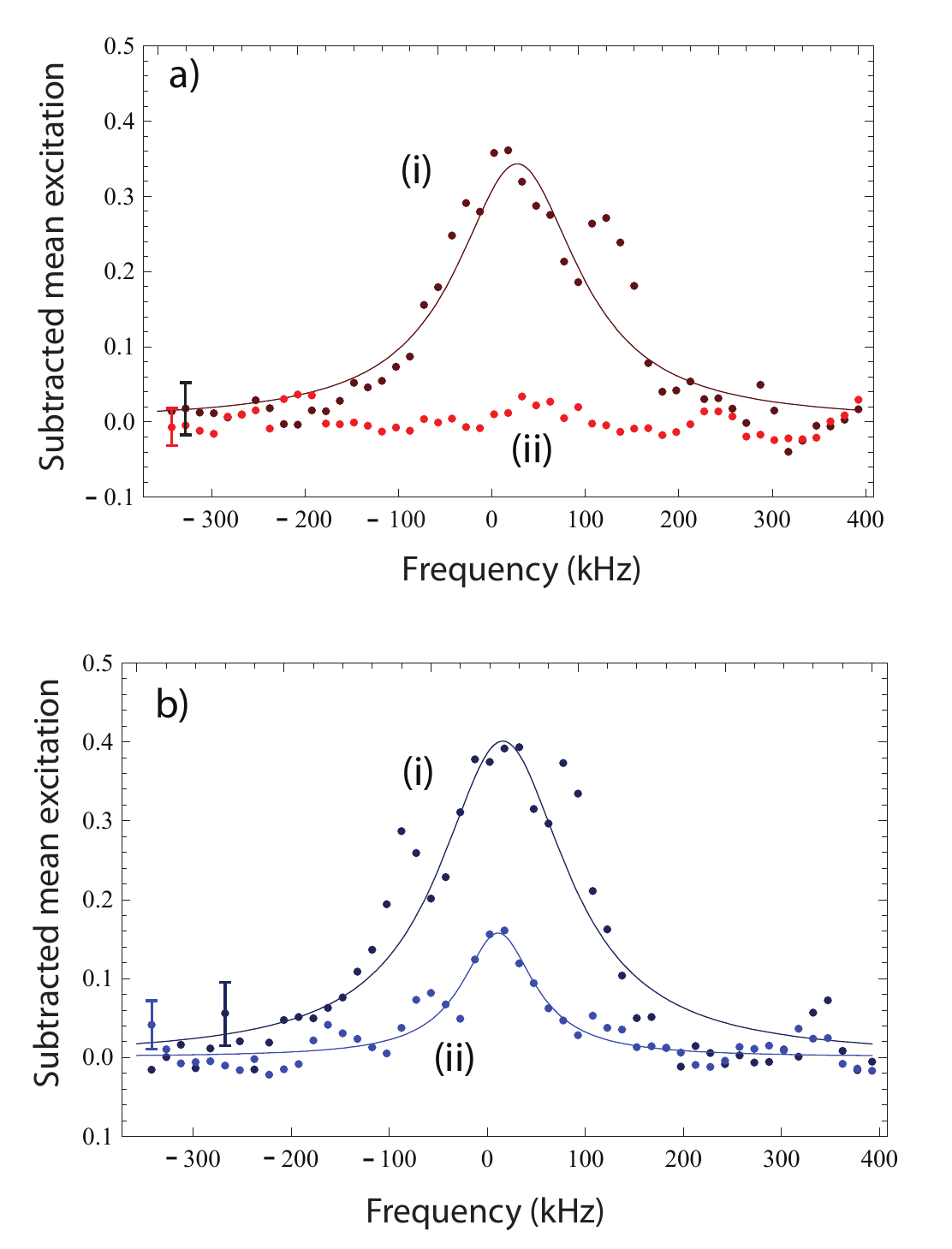}}}
\caption{(Color Online) a) Mean excitation of the a) red and b) blue axial sideband (i) before and (ii) after sideband
cooling. The mean excitations on the sidebands are obtained after subtracting the off-resonant carrier contribution. Error bars show one standard deviation coming mostly from projection noise, averaged over all laser detunings.} \label{SCooling}
\end{figure}

Fig.~\ref{SCooling} shows scans of the mean excitation across the red
and blue axial motional modes.
Fig.~\ref{SCooling}-a (and b) shows the red (and blue) axial sideband (i) before and (ii) after sideband cooling for 10 ms.
To plot these graphs we had to subtract the contribution from the off-resonant carrier excitation. The carrier is indeed only 0.85 MHz away from the sideband and contributes to the signal at the laser excitations that we have to use for estimating precisely the sideband amplitude. The negative values for the mean excitation probability in Fig. 5 result from the subtraction of the carrier background that we perform in order to estimate the axial excitation probability and are in agreement with the excitation probability's measurement error.
The clear reduction of the area under the red sideband shows that the mean photon number has been greatly reduced. The mean phonon number $\langle n\rangle$ and probability $P_0$ for the ion to be in the motional ground state can be estimated from the ratio $R$
between the red and blue sideband areas as
$\langle n\rangle = R/(1 - R)$ and $P_0 = 1 - R$. We reached $\langle
n\rangle=0.03\pm 0.14$ which gives $P_0=97$\,\%$\pm 0.14\%$.
The cooling speed and efficiency are limited mainly by the small
Lamb-Dicke factor. Increasing the 1.76~$\mu m$ laser Rabi frequency helps
only up to the point where sideband cooling excitations are
competing with heating processes coming from off-resonant
carrier excitation. Sideband cooling close to the
motional ground-state was also performed on both radial motional
modes. Since the two radial modes are non-degenerate, we either have to set the frequency of the 1.76~$\mu$m in between the two motional modes or to sequentially cool them by changing the laser frequency. The last option means a longer cooling sequence and a more demanding compensation for AC-Stark shifts. Using the first option, we could nonetheless reach mean phonon numbers of $\langle n\rangle=0.76 \pm 0.11$ for the sum of the two radial modes.

\section{Interferometric thermometry}

Several methods may be employed for measuring single atom temperatures \cite{Lei96,Roo99,Nor11,Knu12}. Here, we make use of the interferometric set-up \cite{Esc01} that is depicted Fig.~\ref{setup1}. Using the quadrupole transition for sideband cooling of the ion and for estimating the mean phonon number allows us to study the self-interference contrast of the ion fluorescence for very low ion temperatures.

Any motion of the ion modulates the phase of the emitted light
which, in this set-up, results in the modulation of the light intensity detected on the PMT. Considering the trapped ion as a quantum harmonic oscillator \cite{Esc03}, the detected intensity will then be given by
\begin{eqnarray}\label{IntensityPh}
I(t) &=& I_0 (1 + V_0 \sum_{n_x,n_y} P_{n_x} P_{n_y} \langle n_x,n_y |\cos[2kL \\\nonumber
&+& 2k(\hat{x}\cos\theta + \hat{y} \cos\theta'+A_m \cos(\Omega_{RF} t))]| n_x,n_y\rangle),
\end{eqnarray}
where $I_0$ is the mean fluorescence rate in counts/sec, $L$ is the ion-mirror distance, $V_0$ is
the visibility of the interference fringes due to imperfect optics
and acoustic noise on the mirror path, $k$ is here the wavevector of the 493\,nm light.
$\hat{x}$, $\hat{y}$ indicate the position operators of
the ion along the respective trap axes, both of which make an angle $\theta=\theta'=\pi/4$ with respect to the mirror-detector direction \cite{Ger09}. Note that we neglected the contribution of the axial motion to the reduction of interference contrast because it is perpendicular to the detector-mirror axis. $P_n=\overline{n}^n/(\overline{n}+1)^{n+1}$ is the thermal distribution with mean number of phonon $\overline{n}$. $A_m$ and $\Omega_{RF}$ are amplitude and frequency of the residual micromotion along the detector-mirror axis. The change of the macromotion amplitude due to the inherent micromotion is on the order of $q \sigma$, where $q$ is the stability parameter, and $\sigma$ is the standard deviation of the ion's position. It is defined as
\begin{eqnarray}
\sigma^2 &=& (2\overline{n}+1)\langle 0 | x_M^2 |0 \rangle.
\end{eqnarray}
 $q$ is $\approx 0.15$ for our trap,
 so this effect is always negligibly small compared to $\sigma$ and was neglected here. The micromotion that we consider here is due to unwanted dielectric patches that displace the ion from the radio-frequency null. Although it was compensated for to a large extent by minimizing the correlation signal between scattered light and the radio-frequency drive \cite{Ber98} and by further optimizing the fringe contrast, it can still increase the motional amplitude by a few nanometers.

\begin{figure}[!ht!]
\centerline{\scalebox{0.8}{\includegraphics{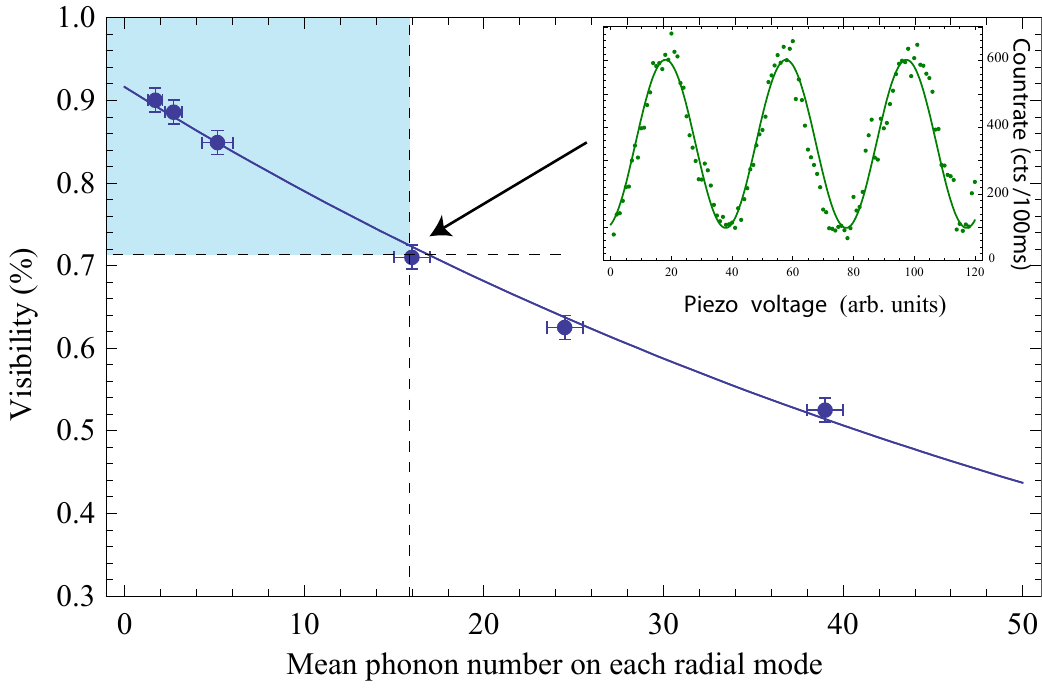}}}
\caption{(Color Online) Dependence of the visibility on the mean phonon
number in the radial modes. The errorbars correspond to one standard deviation. The solid line is a theoretical fit using Eq.\ref{fringesExp}. The inset shows the fluorescence intensity as a function of the mirror position for the mean phonon number of 16. The shaded area highlights the points where the ion was cooled below the Doppler limit. Only the top and bottom values of the intensity are measured for estimating the fringe contrast there. }\label{fringes}
\end{figure}

As shown in Fig.\ref{setup1}, a single mode fiber was used to collect both the direct and retro-reflected fields. Optical abberations due to the overall imaging system are then reduced to a large extent. If distortions of the wavefront on the mirror path are too large however, the fringe contrast will be reduced due to power mismatch. To ensure that the retro-reflected field has the same intensity as the direct field, we then lower the direct field coupling efficiency to the fiber by a slight misalignment of the fiber coupler in order to exactly match this abberation-induced losses. We estimated these losses to be $\approx 40\%$.
Realigning the mirror channel through the new spatial mode defined by the fiber thus ensures that the fringe contrast does not depend on abberations, at the cost of a lower total count rate. After this procedure, the possible contributions to the visibility are then micromotion, radial motion of the ion and acoustic noise on the mirror path. The latter is thus the only contribution that we insert into the prefactor $V_0$.

Writing $|n_x,n_y\rangle$ in the spatial eigen-basis of the harmonic oscillators and averaging the coherent micromotion amplitude over the detection period, we found that
the total visibility is
\begin{eqnarray}\label{fringesExp}
V=V_0 J_0(2k A_m)e^{-2(k\sigma)^2},
\end{eqnarray}
where $\sigma$ is in our case 4.6 nm for the radial motional amplitude in
the ground state. $J_0$ is the zero-order Bessel function. We assumed here that the radial modes are degenerate and that $\overline{n}=\overline{n_x}=\overline{n_y}$.
Already the finite extension of the ground state wavepacket contributes to 1.4\% of the fringe contrast.
We now wish to experimentally determine the dependence of the visibility on the ion's motion. A plot of the fringe contrast as a function of the mean phonon
number on each radial mode (which have the same mean energy) is shown in Fig.~\ref{fringes}. For the three experimental points close to the round state (shaded area), the mean phonon number was varied by changing the sideband cooling time. We apply a sequence of pulses for sideband cooling and then probe the fringe contrast using a weak $100~\mu s$-long 493 nm pulse. Close to the ground state, the added mean phonon number was measured to be at most $0.5$ due to this probe pulse. The mean phonon number was evaluated, as in the above section, by comparing the areas of red and blue motional sidebands of the quadrupolar transition spectrum.
This measurement method is however not accurate for large mean phonon numbers, where the ratio of the areas does not differ significantly anymore. The three other experimental points were measured using only Doppler cooling with increasing powers of the 493 nm laser. To estimate the mean phonon number there, we chose to fit Rabi oscillations on the blue radial sidebands \cite{Roo99}.

Fig.~\ref{fringes} shows that a clear increase of the fringe contrast is observed
as the ion gets closer to the ground state and a maximum of 90\% was observed for a mean phonon number of 1.1 on both radial modes. The fringe contrast is seen to depend strongly on the ion's secular motion, so that high resolution ($\pm 1$ mean phonon number) is achievable using this technique. The solid line shows a fit using Eq.\ref{fringesExp} and $V'=V_0 J_0(2k A_m)$ as the only fitting parameter. Using $V'=0.9$, the theory reproduces well the data. The difference between the 99\%
contrast expected for a near-sideband cooled atom and the maximal
observed value of 90\% can thus be explained by residual
micromotion and acoustic noise of the mirror.

The strong dependence of the interference contrast on the ion motional amplitude suggests a
possible application of this method for ion's thermometry. In the ideal case, the
fundamental limit to the measurement of ultra-low atomic temperature is imposed solely by the fluorescence shot noise and the weak laser pulse that is necessary to probe the self-interference contrast. Compared to the measurement of the phonon number using blue sideband excitation, this technique is thus more invasive. It is however much simpler as it does not rely on complicated laser sequences and ultra-narrow laser sources.\\

\section{Conclusion}

In summary, we demonstrated Rabi oscillations and ground state
cooling of a Barium ion using its quadrupolar transition at $1.76~\mu m$. We show,
that interference of the ion's fluorescence can provide a
useful tool for a measurement of its motional state. We demonstrate a resolution close to one mean phonon number of the method with the ion cooled close to its motional ground state. This study may be useful for understanding the influence of ion motion in free-space QED, direct coupling to single atoms using high-numerical aperture lenses and entanglement generation using single-photon interference \cite{Cab99}.\\

This work has been partially supported by the Austrian
Science Fund FWF (SFB FoQuS), by the European Union
(ERC advanced grant CRYTERION) and by the Instit\"ut f\"ur
Quanteninformation GmbH. G. H. acknowledges support
by a Marie Curie Intra-European Action of the European
Union.

\end{document}